\documentclass[amssymb,aps,twocolumn,floats,showpacs]{revtex4-1}
\usepackage{color,graphicx,float}
\usepackage{natbib}

\usepackage{dcolumn}
\usepackage{bm}

\begin{document}



\title{Effective Hamiltonian based Monte Carlo for the BCS to BEC crossover in the attractive Hubbard model}


\author{Kanika Pasrija$^1$, Prabuddha B. Chakraborty$^2$, and Sanjeev Kumar$^1$}

\affiliation{%
$1$ Indian Institute of Science Education and Research Mohali, Sector 81, S.A.S. Nagar, Manauli PO 140306, India. \\
$2$ Indian Statistical Institute, Chennai Centre, SETS Campus, MGR Knowledge City, Taramani, 600113 Chennai, India.
}%

\date{\today}

\pacs{74.20.-z, 74.20.De, 71.10.Fd, 74.62.En} 

\begin{abstract}
We present an effective Hamiltonian based real-space approach for studying the
weak-coupling Bardeen-Cooper-Schrieffer (BCS) to the strong-coupling 
Bose-Einstein condensate (BEC) crossover 
in the two-dimensional attractive Hubbard model at finite temperatures. 
We introduce and justify an effective classical Hamiltonian to describe the thermal fluctuations of the relevant auxiliary fields.
Our results for $T_c$ and phase diagrams compare very well with those obtained from more sophisticated and {\it cpu}-intensive numerical methods.
We demonstrate that the method works in the presence of disorder and is useful for a real-space description
of the effect of disorder on superconductivity. 
From a combined analysis of the superconducting order parameter, the distribution of auxiliary fields and the quasiparticle density of states, we 
identify the regions of metallic, insulating, superconducting and pseudogapped behavior. 
Our finding of the importance of phase fluctuations for the pseudogap behavior is consistent with the conclusions drawn from recent
experiments on NbN superconductors. The method can be generalized to study superconductors with non-trivial order parameter symmetries by identifying
the relevant auxiliary variables.
\end{abstract}

\maketitle

\section{Introduction}

The attractive Hubbard model (AHM) is the standard phenomenological model that describes the transition from a high-temperature
metallic or insulating state to a low-temperature superconducting state \cite{Micnas1990,Gyorffy1991}. While the microscopic description requires 
an explanation for the origin of the effective attraction between electrons \cite{Tsuji2011,Hirsch1985a}, the nature of the thermally driven transition can be 
understood within the attractive Hubbard framework. Furthermore, the desired symmetry of the superconducting order parameter can be
realized by appropriate choice of the 
attractive interactions, e.g., an on-site attraction leads to s-wave pairing, a nearest-neighbor (nn) attraction gives rise to 
d-wave pairing, and a next to nn attraction in a two orbital model can describe s$^{+-}$ and s$^{++}$ symmetry \cite{Wang1995, Su2001, Yao2012}.
More recently, the AHM has also been used to identify topological quantum phase transitions \cite{Arikawa2010}.

In the limit of weak coupling, the AHM can be studied within the BCS mean field theory, 
and it provides a complete understanding 
of the thermally driven transition, and
an accurate prediction for the transition temperatures. 
The mean-field theory, however, fails in the strong coupling limit where the transition is
controlled by phase fluctuations. Indeed, an effective $XY$ model for phase fluctuations is used to describe the physics of the
strong coupling superconductivity \cite{Erez2013, Erez2013a}. This strong coupling limit is also known as the BEC limit where 
the superconducting phase is understood as a condensate of pre-formed Cooper pairs.
To describe the crossover from the weak coupling BCS to the strong coupling BEC limit within a single framework is 
a challenging problem. Various state of the art methods have been employed to gain insight into the 
behavior of the superconductor across the BCS to BEC crossover \cite{Kaneko2014,Fuchs2004,Freericks1993,Burovski2008,Simonucci2014,
Spuntarelli2007,Zhao2006,Scalettar1989,Sakumichi2014,Allen2001,Dupuis2004}.
The problem becomes even more challenging in the presence of impurities, which are always present in materials \cite{Pan2001,Cren2001,Dubi2007}. In fact, 
disorder as a control parameter has become a powerful concept in understanding some
fundamental aspects of superconductivity. Recent discovery of a Higgs mode in disordered NbN superconductors is one 
prominent example \cite{Sherman2015}.
Intermediate coupling strength demands for a non-perturbative approach, whereas the
presence of disorder calls for an accurate treatment of the spatial correlations. The methods that rely on translational invariance of the Hamiltonian
are not best suited to study the effect of disorder on superconductivity.
Therefore, the importance of explicit real-space approach for the study of disordered interacting fermionic systems has been realized in 
the recent years \cite{Erez2013, Erez2013a, Tarat2015}.

In this paper, we present a conceptually simple and numerically efficient method for a quantitative description of the finite-temperature behavior 
of the AHM. 
The method treats the weak and the strong $U$ regimes on equal footing, and captures the physics of BCS to BEC crossover. 
We make use of the well known analogy of the 
superconducting pairing amplitudes (complex numbers) with $XY$ spins. The parameters of the effective 
model are calculated by analysing the
variations in energy about the mean-field ground state by considering the relevant phase or amplitude fluctuations.
A comparison of $T_c$ estimates with other methods is presented.  
A quantitative description of the amplitude and phase fluctuations allows us to determine their relative importance across the 
BCS to BEC crossover. On the basis of the superconducting order parameter, the quasiparticle density of states and auxiliary field distributions,
we describe the metallic, superconducting, insulating and pseudogapped phases. We find that the pseudogap phase appears close to the insulating phase,
consistent with recent experiments on NbN superconductors.
Finally, we demonstrate that the method works for
disordered Hamiltonian, and discuss the possible extension to superconductors with non-trivial order-parameter symmetries.

The remainder of the paper is organized as follows: In section II we discuss the model and motivate the method. Section III begins with 
a detailed justification of the effective Hamiltonian method. This is followed by the presentation and discussions of results obtained via 
Monte Carlo simulations within the effective Hamiltonian approach. Section III ends with a demonstration of the applicability of the method for disordered Hamiltonians.
We conclude in section IV.

\section{Model and Method}

We consider an AHM on a two-dimensional (2D) square lattice, given by,

\begin{eqnarray}
H &=& -t\sum_{ \langle ij \rangle, \sigma} [c^{\dagger}_{i\sigma} c^{}_{j\sigma} + H.c.] - U \sum_{i} n_{i\uparrow} n_{i\downarrow} -\mu \sum_i n_i, \nonumber \\
& & 
\end{eqnarray}
\noindent
where, the $c^{\dagger}_{i \sigma}$ and $c^{}_{i \sigma}$ are the fermionic creation and annihilation operators. The interaction between fermions is 
considered attractive, as specified by the negative sign 
in front of the $U$ term in the Hamiltonian. The hopping parameter $t$ defines the 
basic energy scale in the model, and therefore we set $t=1$. $\mu$ is the chemical potential which controls the average electron density in the system.
For all the results presented in this paper we adjust $\mu$ so as to obtain and average filling of $\langle n \rangle = 0.8 \pm 0.01$ electrons per site.

For a mean-field treatment of this Hamiltonian, one proceeds by decoupling the interaction term in the pairing channel leading to the
well known Bogoliubov-deGennes (BdG) Hamiltonian,

\begin{eqnarray}
H_{\rm{BdG}} &=& -t\sum_{ \langle ij \rangle, \sigma} [c^{\dagger}_{i\sigma} c^{}_{j\sigma} + H.c.] - \mu \sum_i n_i  \nonumber \\
 & & -U \sum_{i} [\Delta_{i} c^{\dagger}_{i\uparrow} c^{\dagger}_{i\downarrow} + H.c.],
\end{eqnarray}
\noindent 
where $\Delta_i = \langle c_{i \downarrow} c_{i \uparrow}\rangle$ denote the local pairing amplitudes, which are complex numbers. 
The mean-field solution corresponds to the self-consistent 
values for the local variables $\Delta_i$. Note that we do not absorb $U$ is the definition of $\Delta_i$. 
In the absence of impurities one can proceed by assuming a homogeneous solution for $\Delta_i$, and the model can then be solved analytically by making 
use of the Bogoliubov transformations.
In general, one proceeds by numerically diagonalizing $H_{\rm{BdG}}$ and solving for $\{ \Delta_i \}$ self-consistently, 
without any a-priori restrictions on them.
The BdG mean-field method correctly captures the BCS solution in the weak coupling limit, and describes the transition temperature
and the superconducting gap accurately. However, in the strong $U$ limit it 
severely overestimates the superconducting transition
temperature ($T_c$). It is well known that in the strong coupling limit the superconducting order at low temperature can be understood as a 
BEC of pre-formed cooper pairs. 
Therefore, an effective phase-only model is commonly used to describe the strong coupling limit \cite{Carlson1999}. 
In order to capture the weak to strong coupling crossover, one needs to go beyond the BdG mean-field scheme.

Quantum Monte-Carlo, which is sign-problem free for the AHM, clearly provides a very accurate way to study the model at
arbitrary coupling strength. However, the method is computationally intensive. In the determinantal QMC algorithm using the 
Suzuki-Trotter decomposition, for example, the simulation scales as $N^{3}L$, where $N$ is the size of the spatial lattice and $L$ is the size of 
the lattice in the (Matsubara) time direction. Typical lattice sizes that can be studied using QMC 
are $18 \times 18$ sites \cite{Paiva2004}.
Therefore, it becomes difficult to analyze effects of
disorder on superconductivity using QMC. Another approach that has been proposed for studying models of
superconductivity with quenched disorder, is the static-auxiliary-field Monte Carlo (SAF-MC) \cite{Tarat2015}. 
This is a static version of the QMC where the temporal
dependence of the auxiliary fields is ignored, and only the spatial dependence is retained. This method reduces to the BdG mean-field
method at $T=0$, however, it captures the fluctuations in both amplitude and phase of the superconducting order parameters $\Delta_i$ and
therefore captures the finite-temperature physics of a superconductor at arbitrary interaction strengths. The computational time
for this method scales as $N^4$, and therefore one is still severely limited in terms of accessible lattice sizes. Therefore, further approximations
are commonly used to achieve larger sizes \cite{Kumar2006,Covaci2010,Mukherjee2015}. 

Here, we propose that an effective classical Hamiltonian $H_{\rm{cl}}$ can be used to generate configuration for the complex auxiliary field $\Delta_i$. These
configurations can be generated numerically using the standard importance sampling with Metropolis algorithm. 
Our proposed classical Hamiltonian is given by, 
\begin{eqnarray}
H_{\rm{cl}} &=& H_{\rm{phase}} + H_{\rm{amp}}, \nonumber \\
H_{\rm{phase}} &=& -\sum_{ij} J_{ij}(T) \cos(\phi_i-\phi_j), \nonumber \\ 
H_{\rm{amp}} &=& \sum_i k_i(T) ( \Delta_i - \Delta_0(T) )^2. 
\end{eqnarray}

\noindent
In the above, $\phi_i$, $\phi_j$ denote the phases of the superconducting amplitudes $\Delta_i$ and $\Delta_j$ at sites $i$ and $j$, respectively. 
The temperature-dependent parameter $J_{ij}(T)$ denotes the phase stiffness, which will also be bond dependent in the disordered case. 
The term $H_{it{amp}}$ captures the effect of 
amplitude fluctuations about the mean amplitude value for a given temperature. The amplitude stiffness parameter, $k_i(T)$, is in general dependent on 
site as well as temperature. For the clean case the phase and amplitude stiffness parameters are spatially uniform. We further assume that both these
parameters are also independent of temperature. However, it is very important to retain the temperature-dependence of the $\Delta_0(T)$ as will be discussed later.
Within a semi-classical approach, the physics of the Hamiltonian Eq. (1) can then be described by a combination of $H_{\rm{cl}}$ and $H_{\rm{BdG}}$. 
The $H_{\rm{BdG}}$ describes the response of the fermions to a configuration of classical auxiliary field $\Delta_i$, and the evolution of
the auxiliary field is approximately captured by the classical Hamiltonian $H_{\rm{cl}}$. 
This approach is similar, in spirit, to the methods proposed for describing magnetism in the double-exchange model \cite{Calderon1998,Kumar2005}.
The simulations begin at low temperature and we assume the starting state to be a  
phase coherent superconducting state. The mean field solution is obtained at each temperature. By analysing the nature of fluctuations around the
mean-field solution, as will be discussed in the next section, we define the parameters for the  $H_{\rm{cl}}$. The resulting $H_{\rm{cl}}$ is then simulated via Monte Carlo, 
and electronic properties are obtained by using the configurations for ${\Delta_i}$ into $H_{\rm{BdG}}$. Metropolis algorithm with the standard single-site updates
is employed for simulations. Most of the results presented here are obtained on a square lattice with $N=40^2$ sites. Number of Monte Carlo steps 
used for equilibration and averaging of quantities involving classical auxiliary variables is $\sim 10^5$. Since electronic properties 
require a solution of Schroedinger equation for each configuration, thermal averaging for electronic properties is performed over $\sim 10^2$ Monte Carlo steps.

\section{Results}

\subsection{parameters of the effective Hamiltonian}

We begin by analyzing the nature of fluctuations about the BdG mean-field solution of the AHM. Given the complex 
nature of the variables ${\Delta_i} \equiv \vert \Delta_i \vert e^{i \phi_i}$ we can compute the change
in energy caused by the variation in the phases $\phi_i$ and that caused by the change in amplitudes $\vert {\Delta_i} \vert$. 
In order to provide a simple geometrical picture, the ${\Delta_i}$ can be viewed as two dimensional (2D) rotors of variable length $\vert {\Delta_i} \vert$.
\begin{figure}
\includegraphics[width=.98\columnwidth, angle=0, clip = 'True' ]{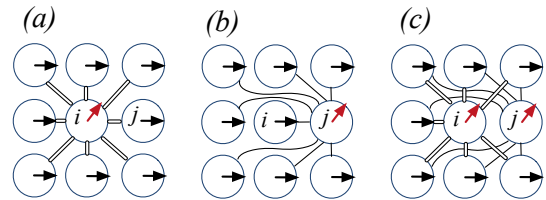}
\caption{ (Color online) A schematic picture describing the method to isolate the contribution of a single rotor-pair to the total energy for the effective classical Hamiltonian. 
($a$) $i^{\rm{th}}$ rotor is oriented away from the otherwise phase coherent arrangement of rotors. The double lines connecting site $i$ to all other sites
indicate the pairs that contribute to the change in energy due to change in the orientation of $i^{\rm{th}}$ rotor.
($b$) $j^{\rm{th}}$ rotor is rotated by an angle $\theta$, and the single lines indicate the pairs contributing to change in energy. 
($c$) Both $i^{\rm{th}}$ and $j^{\rm{th}}$ rotors are rotated by an angle $\theta$. Note that in this case the pair $ij$ does not contribute to the change in energy.
}
\label{fig1}
\end{figure}
It is well known that in the strong coupling limit, $XY$ model captures the physics of phase fluctuations. Moreover, the simplest scalar that 
can be constructed from two vectors is their dot product. 
Therefore, it is reasonable to assume that the change in energy due to relative change of orientation between pairs of rotors is 
described by the first term in $H_{\rm{cl}}$ Eq. (3), {\it i.e.},

$$
H_{\rm{phase}} = - \sum_{ij} J_{ij} \cos (\phi_i - \phi_j),
$$

\noindent
where, in principle, all pairs $ij$ can contribute to the summation.
The task is now to determine the coupling constants $J_{ij}$, which in a translationally invariant system should only depend on the 
distance between sites $i$ and $j$.
Suppose $E_0$ is the energy of the self-consistent BdG solution that in the rotor picture corresponds to all rotors 
pointing in the same direction, say $\phi_i \equiv 0$. 
Now we change the orientation of the rotor at the $i^{\rm{th}}$ site 
by an angle $\theta$ so that $\phi_i = \theta$, and compute the change in energy $\delta E_1$. Within the effective rotor 
model, $H_{\rm{phase}}$, this change must be attributed to the change in bonds that 
connect $i^{\rm{th}}$ site to all other sites (see Fig. \ref{fig1} ($a$)). Next, we restore the orientation of the $i^{\rm{th}}$ rotor back to $\phi_i = 0$,  
and change the orientation of the rotor at the $j^{\rm{th}}$ site by the same angle $\theta$ (see Fig. \ref{fig1} ($b$)). 
This leads to a change in
energy $\delta E_2$ which is coming from the change in bonds that are 
connecting the $j^{\rm{th}}$ site to all other sites. Clearly, for a translationally invariant system we should have $ \delta E_2 = \delta E_1$. 
Then we orient both the $i^{\rm{th}}$ and the $j^{\rm{th}}$ rotors at an angle $\theta$, {\it i.e.}, $\phi_i = \phi_j = \theta$. 
The change is energy obtained in this
configuration is $\delta E_3$. The change in this case is coming from the change in all the bonds connecting $i^{\rm{th}}$ and $j^{\rm{th}}$ rotors to 
all other rotors, except to each other (see Fig. \ref{fig1} ($c$)). Therefore, we can identify the coupling
strength between the $i^{\rm{th}}$ and the $j^{\rm{th}}$ rotors as, $2 J_{ij} = \delta E/(1 - \cos \theta)$ where 
$\delta E = (\delta E_1 + \delta E_2 - \delta E_3)$. Using this protocol for calculating the coupling constants, 
we can also compute the longer-range coupling strengths. 
Note that we are not assuming that only nn bonds contribute to the summation in Eq. (3). In fact, the present scheme for calculating the $J_{ij}$
shows explicitly that the most important coupling is that between the nearest neighbor $i,j$ pairs.
Moreover, the protocol proposed above for computing coupling strengths also works for a disordered system, where $J_{ij}$ will now 
depend on the sites $i$ and $j$, and therefore we will get a distribution of coupling 
strengths even for nearest neighbor couplings.

\begin{figure}
\includegraphics[width=.96\columnwidth,angle=0, clip = 'True' ]{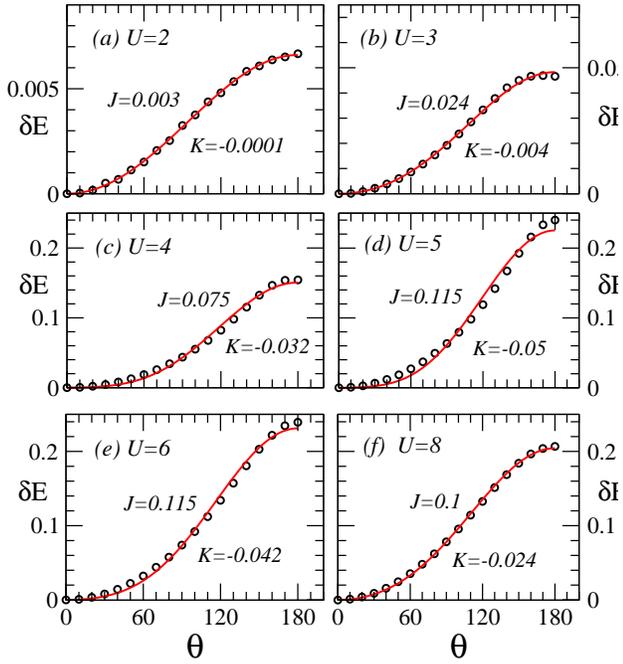}
\caption{(Color online) ($a$)-($f$) Change in energy as a function of the orientation angle between a single pair of rotors for different 
values of $U$. Symbols are the results of numerical calculations and the solid line in each panel is a fit to the
functional form $J(1-\cos \theta) + K(1-\cos^2 \theta)$. The best-fit values of $J$ and $K$ are indicated in the figure.
}
\label{fig2}
\end{figure}

We begin by verifying the validity of the
$\cos(\phi_i-\phi_j)$ form that is assumed in the effective classical Hamiltonian. Fig. \ref{fig2} shows the 
numerical data for change in energy $\delta E$ as a function of angle
of orientation $\theta$ for different values of attractive Hubbard parameter $U$. The function $f({\theta}) = J(1-\cos \theta) + K (1-\cos^2 \theta)$ 
fits the numerical data very well for all values of $U$. The best-fit parameter $J$ is much larger than $K$, therefore in the simplest approximation
we retain only the 
$\cos (\phi_i - \phi_j)$ form in the effective Hamiltonian Eq. (3). In order to compute the values of the coupling parameters $J_{ij}$ one can either use the
best-fit values as indicated in Fig. \ref{fig2} or any two points from the numerical data.

\begin{figure}
\includegraphics[width=.96\columnwidth,angle=0, clip = 'True' ]{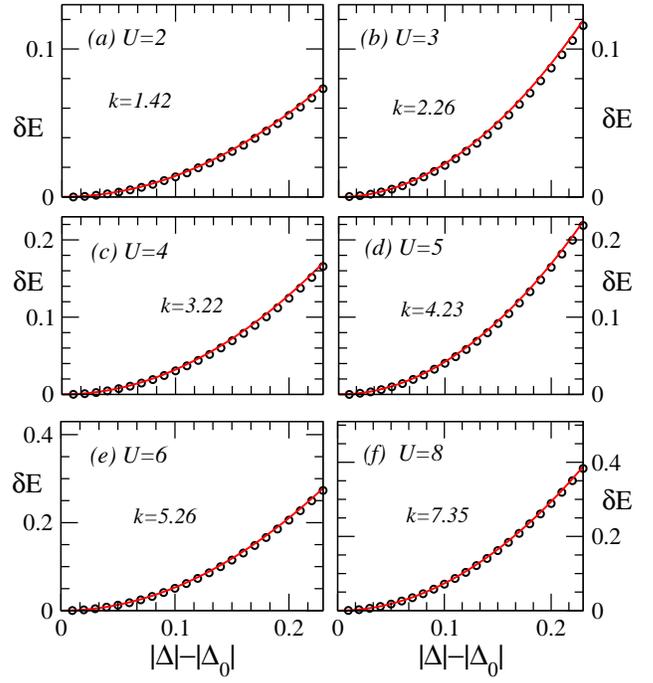}
\caption{(Color online) ($a$)-($f$) Change in energy as a function of the change in magnitude of a single rotor. Symbols are the numerical data and 
the solid line shows a fit to the functional form 
$\delta E = k (\vert \Delta \vert - \vert \Delta_0 \vert)^2$, with the values of best-fit parameter noted in the figure.
}
\label{fig3}
\end{figure}

Following an analogous approach we justify the use of second term in the effective Hamiltonian. This term can be written as
$H_{\rm{amp}} = \sum_i k_i (\vert \Delta_i \vert - \vert \Delta_0\vert )^2$, and represents the stiffness to the change in magnitude of the local pairing amplitude compared to the
average magnitude in the self-consistent solution. Given the on-site nature of this term, it is easier to compute the change in energy.
The results are shown in Fig. \ref{fig3}.
shows the change in energy due to the change in the length of the rotor for different $U$. In this case the function 
$g(\delta \vert \Delta \vert) = k ( \delta \vert \Delta \vert)^2 \equiv k (\vert \Delta \vert - \vert \Delta_0 \vert) ^2 $ 
fits the numerical data very well hence justifying the form of the second term in the effective Hamiltonian.
The rotor picture for the superconducting amplitudes is strictly valid in the large $U$ limit. This is 
analogous to how in the repulsive Hubbard model a local magnetic moment is well-defined only in the large $U$ limit. Therefore, an alternate approach
is used to find the phase stiffness constant in the small $U$ regime. This is obtained from the 
the expectation value of the kinetic energy operator \cite{Toschi2005}.

\begin{figure}
\includegraphics[width=.96\columnwidth,angle=0, clip = 'True' ]{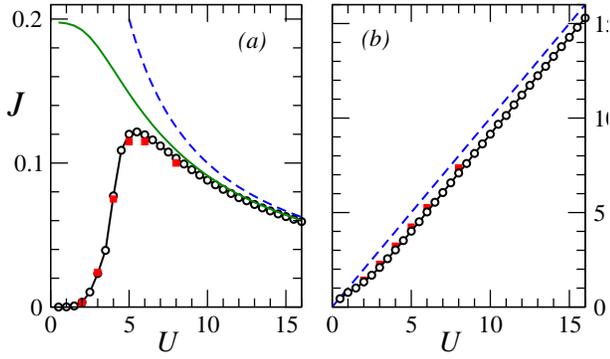}
\caption{(Color online) Parameters of $H_{\rm{cl}}$ as extracted from the change in energy about the mean field value: ($a$) Coupling 
constant $J$ (open circles) as a function of $U$ calculated by assuming a 
cosine form for the change in energy and using only $\theta=0$ and $\theta=\pi$. filled squares show a comparison with $J$ obtained 
from the fits shown in Figure \ref{fig1}. Green-solid line is the phase stiffness calculated as expectation value of the kinetic energy operator.
The dashed-blue line represents the $1/U$ behavior valid in the large $U$ limit.
($b$) The stiffness $k$ to change in local amplitude $\vert \Delta_i \vert$, as a function of $U$ calculated by assuming a $k (\vert \Delta \vert - \vert \Delta_0 \vert)^2$ form and using only 
two points from the data (open circles). Filled-squares are the values
obtained from the best-fits shown in Figure \ref{fig2}. Dashed-blue line represents $k = U$.
}
\label{fig4}
\end{figure}

In the following, we summarize the behavior of the parameters of our $H_{\rm{cl}}$. Fig \ref{fig4}($a$) shows the plot of nn coupling constant $J$ 
as a function of $U$. The values obtained via the best-fit to the cosine form (filled squares in Fig \ref{fig4}($a$)) and those 
obtained by using only $\theta = 0, \pi$ on the cosine curve (open symbols) match very well. 
In the large $U$ limit, we find that $J \sim t^2/U$ as expected from
the strong coupling expansion (dashed line in Fig \ref{fig4}($a$)). 
For $U \leq 5$, $J$ decreases upon decreasing $U$. This indicates a breakdown of the local description for the superconducting amplitudes as 
the phase stiffness at weak coupling should not go to zero in a superconducting phase. Indeed, the phase stiffness computed as the expectation value
of the kinetic energy operator approaches a constant value of around $0.2$. In the intermediate to large $U$ limit the calculations obtained within 
the rotor model are consistent with those obtained in the standard kinetic energy approach \cite{Toschi2005}. 
The amplitude stiffness parameter $k$  as a function of $U$ is shown in Fig \ref{fig4}($b$). Once again using a best-fit to the quadratic form (filled squares) and
using only two points from the numerical data (open symbols) are very close. The dashed line corresponds to $k = U$, and seems to be a good approximation for the
stiffness constant over the entire $U$ range.

\subsection{Monte Carlo simulation results}

\begin{figure}
\includegraphics[width=1.0\columnwidth,angle=0, clip = 'True' ]{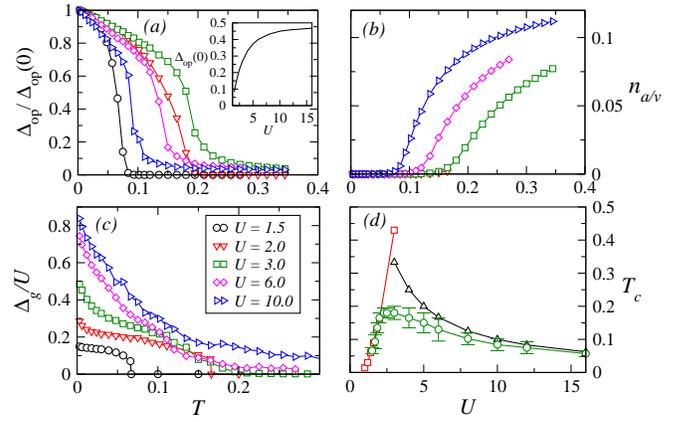}
\caption{(Color online) ($a$) The temperature dependence of the superconducting order parameter $\Delta_{{\rm op}}$ normalized by its low temperature value, $\Delta_{{\rm op}}(0)$, 
for different values of $U$. Inset shows the variation of $\Delta_{{\rm op}}(0)$ with $U$. ($b$) Vortex $n_v$ and anti-vortex $n_a$ density as function of temperature. 
($c$) Spectral gap $\Delta_{g}$ as a function of temperature for different $U$. 
($d$) Circles show the transition temperature $T_{c}$, as inferred from the inflection point in the $T$-dependence of the order parameter, as a function of $U$.
Squares and triangles mark, respectively, the expected variations of $T_c$ in the small-$U$ and large-$U$ limits.
}
\label{fig5}
\end{figure}

We define the superconducting order parameter at finite temperature by, $ \Delta_{{\rm op}} =  \frac{1}{N} \langle \sum_i \Delta_i \rangle$, where the angular
brackets denote thermal averaging over Monte Carlo configurations of auxiliary variables and $N$ is the number of sites.
The temperature dependence of $\Delta_{{\rm op}} $ for different values of $U$ is shown in Fig. \ref{fig5} ($a$). The point of inflection in
$\Delta_{{\rm op}}(T)$ is used to estimate the value of the superconducting transition temperature $T_c$.
The transition temperature displays a non-monotonic behavior with varying $U$ (see Fig. \ref{fig5}($a$)). The sharp reduction in $\Delta_{{\rm op}}$
across $T_c$ is caused by the vanishing of $\vert \Delta_i \vert$ for small $U$, and by randomness in phases $\phi_i$ for large $U$.
These two limits are connected smoothly with variation in $U$, as will be discussed in detail in the following.

A useful quantity that determines the importance of the phase of the superconducting order parameter is
the vorticity \cite{Erez2010}. Vorticity (antivorticity) can be defined as the sum of difference of phases around a square plaquette taken clockwise (anticlockwise)
and summed over all plaquettes. The difference in angles $\phi_j - \phi_i$ is defined modulo $\pm \pi$.
The density of vortices and anti-vortices ($n_{v/a}$) is shown in Fig. \ref{fig5} ($b$).
In the weak coupling regime there are no vortex/antivortex excitations as the system goes across the transition 
(see Fig. \ref{fig5} ($b$) for $U=1.5$ and $U=2$). This shows that the transition is caused
solely by fluctuations in amplitudes of the local superconducting order parameters $\Delta_i$. Indeed, for intermediate to large values of $U$, density of
vortices begins to rise near the transition temperature as determined from $\Delta_{{\rm op}}$.
This is consistent with previous results obtained in the extreme large $U$ limit, where
one can assume the magnitudes $\vert \Delta_i \vert$ to be constant and the fluctuations are captured by a phase-only $XY$ model \cite{Erez2010}.

The electronic spectrum is obtained in the Monte Carlo generated auxiliary field configurations by solving for $H_{\rm{BdG}}$ Eq. (2).
One of the important features contained in the electronic spectra is the spectral gap, which we define as the 
energy difference between lowest unoccupied level and highest occupied level assuming a $T=0$ Fermi distribution function.
The spectral gap normalized to the value of $U$ is plotted in Fig. \ref{fig5} ($c$). The temperature dependence shows that the gap vanishes at
$T_c$ for small values of $U$, whereas it remains finite even in the non-superconducting regime for intermediate to large values of $U$.
The $U$-dependence of $T_c$ obtained in present study is consistent with the BCS result for small $U$ and a strong coupling $1/U$ 
behavior for large $U$ (Fig. \ref{fig5} ($d$)).
These results are qualitatively similar to those obtained by the Dynamical Mean Field Theory (DMFT), QMC and other computationally demanding methods. 
The quantitative features are as follows.
The maximum value of $T_c$ is $0.18 t$, and occurs near $U=3.0 t$. Within various methods these characteristic scales are respectively given by,
$0.12 t$ and $2t$ (T-matrix approximation), $0.2t$ and $4t$ (DMFT), $0.16 t$ and $4t$ (fluctuation exchange approximation), $0.18 t$ and $5t$ (QMC), and
$0.14 t$ and $5t$ (SAF-MC) \cite{Keller2001,Deisz2002,Paiva2010,Tarat2015}. Although the maximum value of $T_c$ and the corresponding $U$ value should both depend 
on the average electron density, 
within QMC this dependence is insignificant in the density range $0.5 < n < 0.9$, and hence the above comparison is meaningful 
despite the different values on $n$ used in different studies \cite{Paiva2004}.

\begin{figure}
\includegraphics[width=1.0\columnwidth,angle=0, clip = 'True' ]{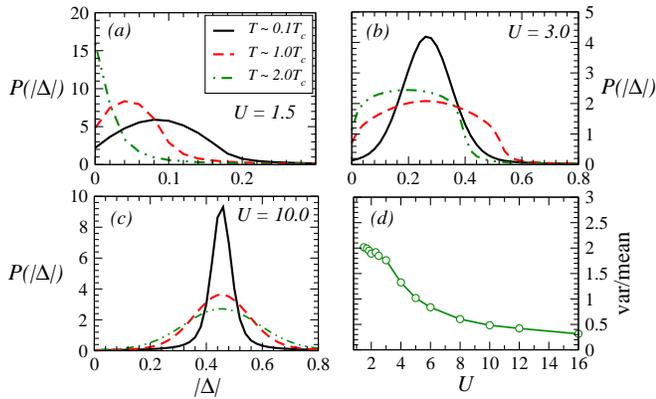}
\caption{(Color online) The distribution $P( \vert \Delta \vert)$ of magnitudes $\vert \Delta \vert$ at different temperatures, $0.1T_{c}$, $1.0T_{c}$, $2.0T_{c}$, for    
($a$) $U=1.5$, ($b$) $U=3$ and ($c$) $U=10$. ($d$) The ratio of the variance to the mean value of $P(\vert \Delta \vert)$ as function of $U$. 
This ratio decreases with increasing $U$ highlighting the importance of amplitude fluctuations at weak $U$ values.}
\label{fig6}
\end{figure}

The introduction of an effective classical Hamiltonian for auxiliary fields has two-fold advantage. 
Firstly, it facilitates the application of the Monte Carlo procedure, secondly, the behavior of auxiliary variables
provides additional insight into the nature of the finite-temperature transitions.
In order to further understand the difference between the superconducting to normal state transitions at weak and strong coupling, we investigate the details of the
temperature evolution of local pairing amplitudes, $\Delta_i$. The distribution of the magnitude of pairing amplitudes is computed via,

\begin{equation}
P(\vert \Delta \vert) = \frac{1}{N} \left \langle \sum_i \delta( \vert \Delta \vert  - \vert \Delta_i \vert ) \right \rangle,
\end{equation}

\noindent
where, the Dirac-delta function is approximated by a Lorentzian with width $\eta = 0.01$.
The resulting distribution is plotted in Fig. \ref{fig6} ($a$)-($c$). 
At low $T$ the mean value of the distribution increases with increasing $U$. The width of the distribution decreases with increasing temperature for $U=1.5$ 
due to a decrease in $\Delta_0$ with increasing $T$ (see Fig. \ref{fig6} ($a$)). At large $U$, since $\Delta_0$ becomes almost independent of $T$, 
an expected increase in the width of the distribution due to thermal effects is obtained in our simulations (see Fig. \ref{fig6} ($c$)). 
Interestingly, a combination of these two effects occurs at intermediate $U$ where the width first increases and then decreases upon increasing $T$ (see Fig. \ref{fig6} ($b$)).
In order to assess the relative importance 
of the amplitude fluctuations in driving the system to a normal state, we compute the ratio of the variance to the mean value of the distribution.
This is plotted as a function of $U$ for $T \sim T_c$ in Fig. \ref{fig6} ($d$). Clearly, the amplitude fluctuations become less important upon increasing the strength of
attractive coupling. Nevertheless, such fluctuations are always present, and seem to vanish only asymptotically.

Next, we discuss the fluctuations in the phase of the superconductor along the same lines as those in amplitudes. We define a bond-variable $D_{ij} = \cos (\phi_i - \phi_j)$, where $i$ and $j$ are the nn sites,
and compute the distribution of $D_{ij}$ as,
\begin{equation}
P(D) = \frac{1}{N} \left \langle \sum_i \delta(D - D_{ij}) \right \rangle.
\end{equation}

\begin{figure}
\includegraphics[width=1.0\columnwidth,angle=0, clip = 'True' ]{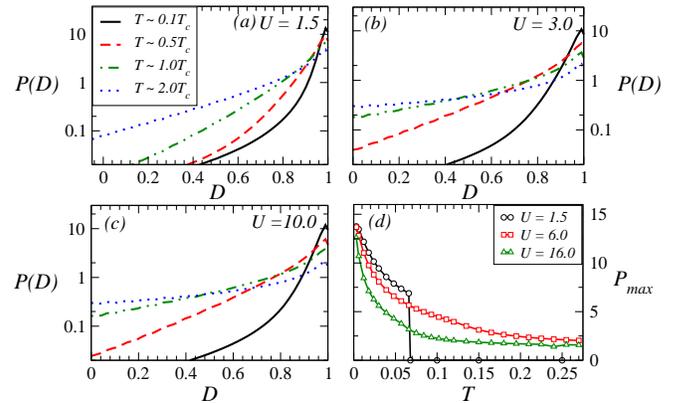}
\caption{(Color online) The distribution $P(D)$ of the nn phase correlators $cos(\phi_{i}-\phi_{j})$ (see text)
at different temperatures, $0.1T_{c}$, $0.5T_{c}$, $1.0T_{c}$, $2.0T_{c}$, 
for  ($a$) $U=1.5$, ($b$) $U=3$ and ($c$) $U=10$. ($d$) The maximum value, $P_{max}$, of $P(D)$ as a function of T for various $U$ values.
}
\label{fig7}
\end{figure}

\noindent
The $\delta$ function is approximated by a Lorentzian as before. The distributions are shown in Fig. \ref{fig7} ($a$)-($c$) for different values of $U$ and $T$. For all values of $U$, the distribution is
sharply peaked near $D = 1$ at low temperatures, and becomes progressively broader with increasing temperature. The inverse of peak-height of the distribution can be taken as an indicator for the 
width of the distribution. In Fig. \ref{fig7} ($d$) we show the peak-height as a function of $T$ for three values of $U$. For intermediate and large $U$, the peak-height reduces strongly with temperature,
indicating stronger fluctuations in the phase. The results are, therefore, consistent with the well known notion that for strong interactions the phase fluctuations are dominant. 
The overall behavior of amplitude and phase fluctuations shows that for a wide intermediate range of $U$, both the amplitude and phase fluctuations play important role in
driving the superconducting state towards a normal state.

\begin{figure}
\includegraphics[width=.92\columnwidth,angle=0, clip = 'True']{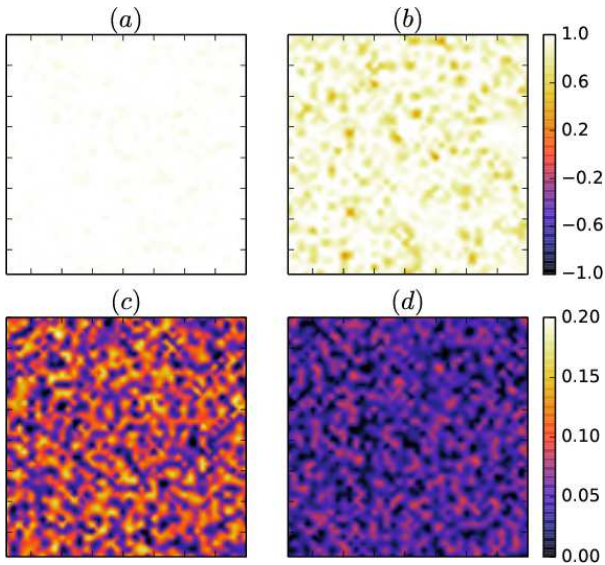}
\caption{(Color online) Real space plots for two different temperatures $T \sim 0.1T_{c}$ (first column) and $T \sim T_{c}$ (second column) for $U/t$ = 1.5.
($a$)-($b$) show the nn phase correlations $D_{ij} = \cos (\phi_i-\phi_j)$ and ($c$)-($d$) the amplitude variables $\vert \Delta_i \vert$. 
}
\label{fig8}
\end{figure}

\begin{figure}
\includegraphics[width=.92\columnwidth,angle=0, clip = 'True']{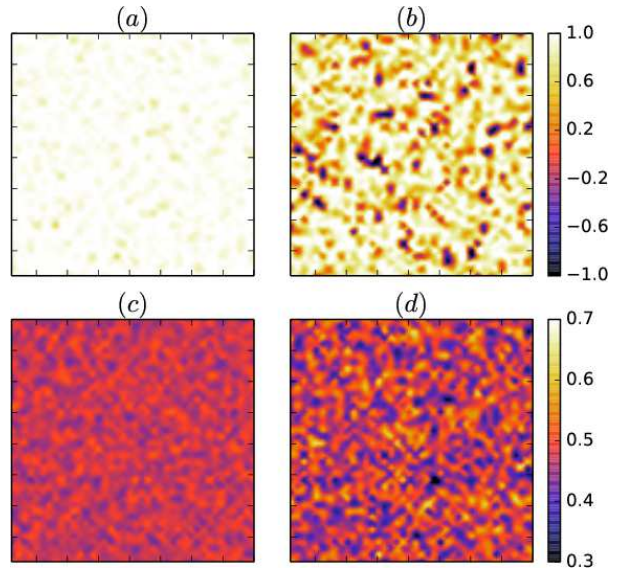}
\caption{(Color online) Real space plots for two different temperatures $T \sim 0.1T_{c}$ (first column) and $T \sim T_{c}$ (second column) for $U/t$ = 16.
($a$)-($b$) show the nn phase correlations $D_{ij} = \cos (\phi_i-\phi_j)$ and ($c$)-($d$) the amplitude variables $\vert \Delta_i \vert$.  }
\label{fig9}
\end{figure}

We plot the configurations of the auxiliary variables in terms of the amplitude and the phase of ${\Delta_i}$. The plot is shown in Fig. \ref{fig8} for $U=1.5$ 
and in Fig. \ref{fig9} for $U=16$ at $T \sim 0.1 T_c$
and $T \sim T_c$. For small $U$ the fluctuations in the phase $\{ \phi_i \}$ are essentially absent at $T \sim 0.1 T_c$, and remain insignificant 
even as $T$ approaches $T_c$ (see Fig. \ref{fig8} ($a$)-($b$)). On the other hand, the amplitudes $\{ \vert \Delta_i \vert \}$ show significant fluctuations 
already at $T \sim 0.1 T_c$, which become very strong as $T$ approaches $T_c$ (see Fig. \ref{fig8} ($c$)-($d$)). This reconfirms that the 
small $U$ regime is dominated by amplitude fluctuations. The trends are essentially reversed for large $U$. The phase fluctuations are relatively stronger
for $U=16$ (see Fig. \ref{fig9} ($a$)-($b$)). The amplitudes also contain significant fluctuations, but remain finite even at $T_c$ (see Fig. \ref{fig9} ($c$)-($d$)). 
Therefore the loss of superconductivity in the large $U$ limit is driven by the fluctuations in the phase.
While the dominant fluctuations can be identified as amplitude-like for weak $U$ and phase-like for strong $U$, fluctuations in both the phase and amplitude 
variables are present over the full range of the attraction strength. The idealized amplitude-only and phase-only descriptions of 
the suppression of superconducting order seem to be valid only in a very small $U$ and very large $U$ regimes of the model.
This is supported by experiments where presence of Josephson effect, which is an indicator for phase-sensitive superconductivity,
is observed over the entire BCS to BEC crossover region \cite{Spuntarelli2007}.

\begin{figure}
\includegraphics[width=1.0\columnwidth,angle=0, clip = 'True' ]{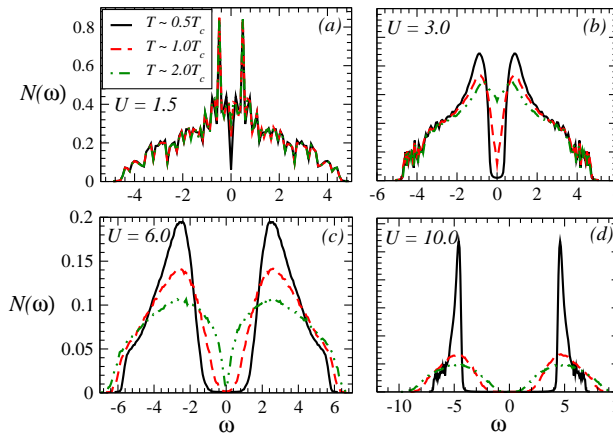}
\caption{(Color online) Variation of quasi particle density of states, $N(\omega)$, with temperatures for coupling strengths ($a$) $U=1.5$, ($b$) $U=3$, ($c$) $U=6$, 
and ($d$) $U=10$. For small $U$, the spectral gap vanishes at $T_{c}$ as expected in the BCS regime.}
\label{fig10}
\end{figure}

The behavior of the classical auxiliary variables with increasing temperature and for different $U$ already provides us with substantial insight into the 
thermal physics of the superconductor. It is equally important to analyze the response of the quasiparticles to the thermal fluctuations at different values of $U$.
To this end, we now discuss the behavior of the quasiparticle density of states, which is defined as,

\begin{equation}
N(\omega) = \frac{1}{N} \left \langle  \sum_k \delta(\omega - \epsilon_k) \right \rangle,
\end{equation}

\noindent
where, $\epsilon_k$ are the $2N$ eigenvalues obtained numerically by solving for $H_{{\rm BdG}}$ equation (2) in a given configuration of the classical auxiliary variables. 
The angular bracket denotes averaging over various ${\Delta_i}$ configurations as generated by the Monte-Carlo. The DOS across the entire range of $T$ and $U$ can be grouped into 
three qualitatively distinct categories based on their behavior near the chemical potential.
These are, (i) gapped, (ii) pseudogapped, and (ii) gapless (see Fig. \ref{fig10}).
At $T=0$ the DOS supports a finite spectral gap for all values of $U$. However, the finite $T$ behavior depends strongly on the value of $U$. For small $U$, the gap
vanishes as $T \geq T_c$ (see Fig. \ref{fig10} ($a$)). For very large $U$ the gap persists above $T_c$ (see Fig. \ref{fig10} ($c$)-($d$)).
In the intermediate to strong $U$ regime, the DOS shows a dip at chemical potential without a clean gap. This regime of parameter space is termed as pseudogap regime.

In order to find a possible connection between the nature of the DOS as discussed above, and the nature of fluctuations in the
auxiliary field variables we consider the following three idealized configurations of auxiliary variables. These are, (i) amplitude only fluctuations: configurations with
perfect phase coherence ($\phi_i \equiv \phi_0$), but a random distribution of $\vert \Delta_i \vert$ between $0$ and $2 \vert \Delta_0 \vert$, where $\Delta_0$
is the low temperature value of the order parameter, (ii) phase only fluctuations: the amplitudes are uniform ($\vert \Delta_i \vert \equiv \vert \Delta_0 \vert$) 
and the phases are randomly distributed between $0$ and $2\pi$, and (iii) amplitude and phase fluctuations: both $\vert \Delta_i \vert$ and $\phi_i$ are randomly
distributed over the above mentioned range. The DOS is computed for these three idealized configurations for different values of $U$. The outcome of this in terms 
on the nature of DOS is presented in Table I.
\begin{table}
\begin{tabular}{|c|c|c|c|}
\hline $\text{$U$} \Downarrow$ ,  $\text{Fluctuations}\Rightarrow$  &      
$ \{ \vert \Delta_i \vert \}$ &      $\{ \vert \Delta_i \vert  \} + \{\phi_i\}$ &    $\{ \phi_i \} $ \\ 

\hline 1.5 & Gapless & Gapless  & Gapless  \\ 
\hline 2.0 & Gapless  & Gapless  & Gapless  \\ 
\hline 3.0 & Gapped & Pseudogapped & Pseudogapped \\ 
\hline 4.0 & Gapped & Pseudogapped & Pseudogapped \\ 
\hline 6.0 & Gapped & Gapped & Gapped \\ 
\hline 8.0 & Gapped & Gapped & Gapped \\ 
\hline 16.0 & Gapped & Gapped & Gapped \\ 
\hline 
\end{tabular} 
\caption{Nature of Density of states (DOS) at $T>T_{c}$ for different values of $U$ within three basic scenarios that consider different combinations of
fluctuations in magnitude and phase of $\Delta_i$.}
\end{table}
We find that amplitude-only fluctuations do not lead to a pseudogapped DOS. For the other two combinations, the pseudogap phase occurs for intermediate
values of $U$, and a fully gapped DOS above $T_c$ is consistent with both phase-only and amplitude and phase fluctuations. This suggests that presence of 
a pseudogap phase can be considered as an indicator for the presence of phase fluctuations. Recent experiments indeed show that a pseudogap phase can
exist in conventional superconductors that sit at the proximity to an insulating phase \cite{Sacepe2010}.

\begin{figure}
\includegraphics[width=1.0\columnwidth,angle=0, clip = 'True' ]{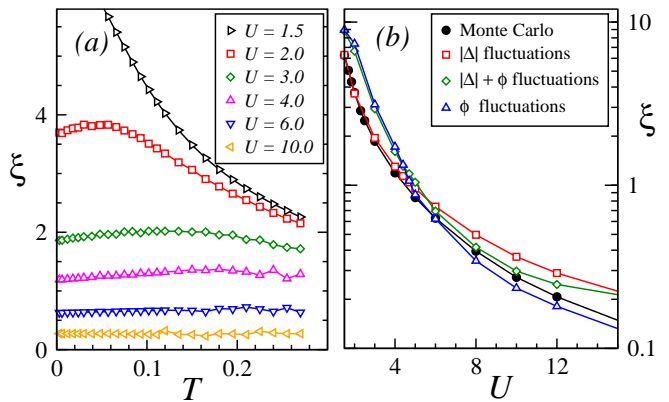}
\caption{(Color online)($a$) Temperature variation of coherence length $(\xi)$ for different $U$. ($b$) variation of $\xi$ as function of $U$ at low temperatures.
The filled symbols represent the Monte Carlo data, and the other three data sets are for the hypothetical auxiliary field configurations corresponding to 
fluctuations in magnitude of $\Delta_i$, phase of $\Delta_i$, and both phase and magnitude of $\Delta_i$ (see text). The results for coherence length are
obtained on $64 \times 64$ lattice.
}
\label{fig11}
\end{figure}

Another indicator that is commonly used to describe the crossover from a BCS-like superconductor to the BEC of cooper pairs is the coherence length
of the superconductor. The coherence length $\xi$ is defined via \cite{Kaneko2014},

\begin{equation}
\xi^{2}= \frac{ \sum_{r} r^{2}\vert F(r)\vert ^{2}}{\sum_{r} \vert F(r)\vert ^{2}},
\end{equation}

\noindent
where $F(r) = \frac{1}{\sqrt{N}} \sum_{i} \langle c_{i+ r \downarrow} c_{i\uparrow} \rangle$, and
$i+r$ denotes a site located at distance $r$ from site $i$.
Fig. \ref{fig11}($a$) shows the temperature dependence of pair coherence length for different values of $U$.
For small values of $U$, the coherence length decreases with temperature, and $\xi(T_c)/\xi(0) \sim 0.8$ in agreement with previous calculations \cite{Kaneko2014}.
With increasing $U$, $\xi$ reduces rapidly and becomes essentially temperature-independent. Note that $\xi < 1$ for $U > 4$ indicates that the cooper pairs have
essentially become well localized in this regime of interaction strength.
We further test the three basic scenarios of fluctuations in auxiliary variables for the pair coherence length.
We compare the results obtained for the pair coherence length in the Monte Carlo simulations, with those obtained
by considering three types of idealized auxiliary variable configurations that are already discussed for the DOS.
We find that for small values of $U$, our Monte Carlo simulation results for $\xi$ are very close to those obtained in the 
amplitude-only fluctuation model (see Fig. \ref{fig11}($b$)). In the large $U$ regime, the Monte Carlo results are closest to the phase-only fluctuation model.
In the intermediate range, $4 < U < 10$, the coherence length is best described by the fluctuations in both ${\vert \Delta_i \vert}$ and ${\phi_i}$
These results indicate that the Monte Carlo method faithfully captures the crossover from amplitude-only fluctuation regime at small $U$ to 
the $XY$-model regime at large $U$.

\begin{figure}
\includegraphics[width=1.0\columnwidth,angle=0, clip = 'True' ]{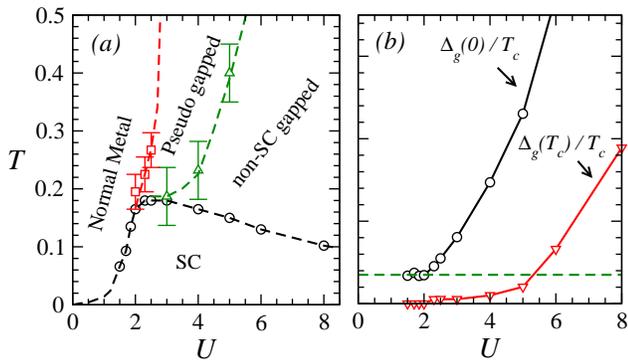}
\caption{(Color online)($a$) The $T$-$U$ phase diagram showing the superconducting, the normal metal, the 
non-superconducting gapped and pseudogapped phases. Symbols are the data points obtained from simulations and the
dashed lines are guide to eye. ($b$) The ratio of spectral gap to the ordering temperature, $\frac{\Delta_g}{k_B T_c}$, as a 
function of $U$. The ratio is shown for the gap at $T=0$ and that at $T \sim T_c$.
}
\label{fig12}
\end{figure}

We summarize the results obtained so far in a phase diagram in Fig. \ref{fig12} ($a$). The $T$-$U$ phase diagram as obtained within our simulations 
consists of four distinct phases, namely, SC, normal metal, non-SC gapped, and pseudogapped.
This is consistent with results obtained via more sophisticated numerical techniques.
The pseudogapped state
can be understood as an indicator for the presence of both the phase and the amplitude fluctuations. The turn-around of the $T_c$ vs. $U$ curve,
which is located close to the BCS behavior within our calculations, can be considered as an indicator for the on-set of significant phase fluctuations.
This is why the region just above $T_c$ shows pseudogapped DOS. As $U$ increases, the phase fluctuations become dominant, however the amplitude fluctuations become
inactive only when $U$ is considerably large. The DOS remains gapped as long as amplitude fluctuations are absent, and at higher $T$ when both amplitude and 
phase are random, a pseudogap phase appears. The pseudogap phase is likely to disappear at a scale proportional to $U$ itself, where the pairing amplitudes
themselves vanish and therefore, the phase of the order parameter cannot be defined.
In Fig. \ref{fig12} ($b$) we show the plot of $\Delta_g (0)/T_c$ as a function of $U$. The plot begins to deviate from the BCS value of $3.5$,
(as indicated by the horizontal dashed line) around $U = 2$. We also 
show $\Delta_g(T_c)/T_c$ as a function of $U$. In the BCS scenario, $\Delta_g(T_c)/T_c =0 $  which we find to hold for $U \leq 3$. These two indicators of BCS behavior
suggest that the deviation from a BCS like superconducting order begins somewhere between $U=2$ and $U=3$. However, there is no critical value of $U$ for which the
behavior deviates from the BCS behavior.

\subsection{Effective Hamiltonian in the presence of quenched disorder}

Although disorder is present to varying degrees in almost all materials, its effect is typically ignored in
the simplest treatment. Indeed, translational invariance is commonly invoked in theories of condensed matter systems.
In the context of superconductors, however, disorder plays a crucial role in providing a better understanding of the 
underlying mechanisms. Indeed, there has been immense interest in studying disordered superconductors, both bulk and thin films,
in recent years \cite{Sacepe2010,Sherman2015}. 
The idea is to use disorder as a control parameter which then provides new insights into the understanding 
of correlated electron physics.
Hence, methods that can treat the effect of disorder accurately become extremely useful. 
This is where the real-space methods hold an edge over the variety of mean-field methods.
Having shown that the real-space method proposed in section II of this paper recovers the physics of
thermal fluctuations in both, the amplitude and the phase of the superconducting order, we now demonstrate that 
the scheme can applied to disordered Hamiltonians. 
In order to proceed, we use the prototype model for disorder and extend our Hamiltonian Eq. (1) by adding a random on-site
energy term. The resulting disordered Hamiltonian is given by,

\begin{eqnarray}
H' &=& H + \sum_i \epsilon_i n_i,
\end{eqnarray}

\begin{figure}
\includegraphics[width=.96\columnwidth,angle=0, clip = 'True' ]{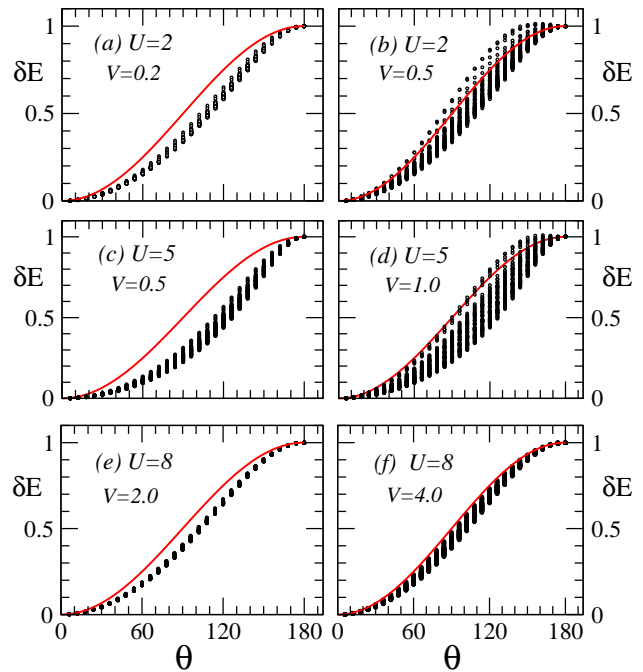}
\caption{(Color online) ($a$)-($f$) Change in energy as a function of the orientation angle between a single pair of rotors for different 
values of $U$ and $V$. The scatter of points is due to the inequivalence of nearest neighbor pairs. In order to show the variation for
different pairs on the same scale, we have normalized the variation in energy for each pair by its maximum value, hence all the points lie between $0$ and $1$.
The solid line in each panel is the function $(1-\cos \theta)/2$.
}
\label{fig13}
\end{figure}

\noindent
where, $\epsilon_i$ are random variables selected from a uniform box distribution of width $V$, {\it i.e.}, $-V/2 < \epsilon_i < V/2$.
The additional term effects both $H_{\rm{BdG}}$ Eq. (2) and $H_{\rm{cl}}$ Eq (3). The change in $H_{\rm{BdG}}$ is simply the addition of the term $\sum_i \epsilon_i n_i$ to Eq. (2).
The change in $H_{\rm{cl}}$ arises via the change in the parameters of the effective Hamiltonian. Since translational symmetry is broken by the 
disorder term, the parameters $J_{ij}$ and $k_i$ in Eq. (3) become site dependent.
However, even before arriving at the effective parameters, we need to verify the validity of the form of the effective Hamiltonian Eq (3).
We show the dependence of the change in energy on the rotation angle for all nn pair of sites. Since our primary task is to identify
the functional form of $\delta E(\theta)$, we plot the change in energies normalized to the change for largest 
value of $\theta$, {\it i.e.}, $\theta = \pi$ for all nn pairs. The resulting plot is shown in Fig. \ref{fig13} for a few representative values
of $U$ and $V$. While there is a broadening due to disorder, the overall shape of the curve is reasonably well approximated by a cosine function.
Interestingly, the deviation from the cosine behavior is large for intermediate values of $U$. For large $V$, the cosine curve passes through the scatter of
points corresponding to $\delta E(\theta)$ for different nn pairs (see Fig. \ref{fig13} ($b$), ($d$), ($f$)). Note that the fit does not appear as good as that in
the clean case (see Fig. \ref{fig2}) because we are not using additional fit parameter, $K$, in this case. In principle, more parameters can be introduced in 
$H_{{\rm cl}}$ in order to improve the model, however, our aim here is to demonstrate the working of the general scheme and therefore we leave this task of
quantitative improvements for future. The results presented for the disordered case are averaged over $4-10$ realizations of disorder.

\begin{figure}
\includegraphics[width=.96\columnwidth,angle=0, clip = 'True' ]{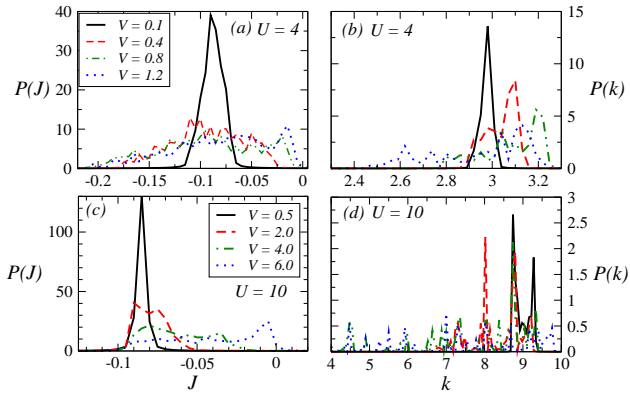}
\caption{(Color online) Probability distributions for coupling (J) for ($a$) $U=4$ , ($c$) $U=10$ and
 probability distributions for stiffness (k) for ($b$) $U=4$ , ($d$) $U=10$ at various disorders.}
\label{fig14}
\end{figure}

\begin{figure}
\includegraphics[width=.96\columnwidth,angle=0, clip = 'True' ]{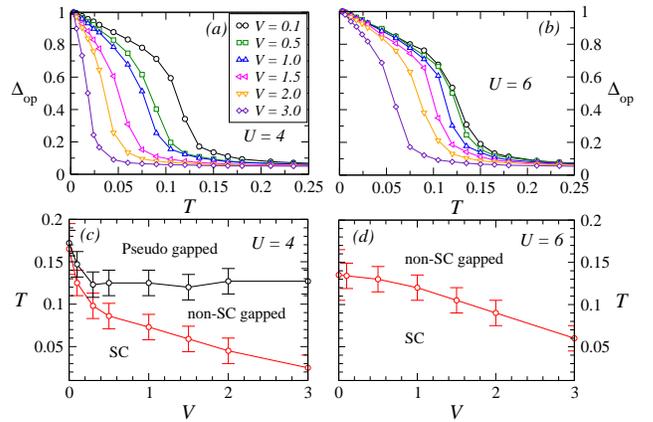}
\caption{(Color online) Temperature dependence of the superconducting order parameter normalized by its low temperature value, $\Delta_{{\rm op}}/\Delta_{{\rm op}}(0)$
for ($a$) $U=4$ and ($b$) $U=6$. 
The phase diagram in temperature-disorder plane 
showing  Superconducting, non-superconducting but gapped, and pseudogapped regimes for ($c$) $U=4$ and ($d$) $U=6$.}
\label{fig15}
\end{figure}

The distribution of parameters is shown in Fig. \ref{fig14}. Both $J_{ij}$ and $k_i$ acquire a broad distribution for finite values of $V$. 
Interestingly, for large $U$ the coupling strengths $J_{ij}$ do not become larger than the disorder-free value of $J$. For the stiffness constant,
$k_i < U$ for all values of disorder strength and $U$, and for all sites. The method employed here for calculating $J_{ij}$ can also be useful in the study of
quantum $XY$ models where the common practice is to select $J_{ij}$ from random uncorrelated distributions \cite{Swanson2014}.
The Monte Carlo simulations proceed as in the case of disorder-free Hamiltonian, except that 
in the present case the parameters $k_i$ and $J_{ij}$ of $H_{\rm{cl}}$ are site and bond dependent, respectively.
From the behavior of the parameters for $H_{{\rm cl}}$ in the presence of disorder, we can already argue that the fluctuations in both the amplitude and the phase
of the superconducting order parameter are enhanced by disorder.
We show the results for the superconducting order parameter in Fig. \ref{fig15} ($a$). The $T=0$ value of the order parameter decreases rapidly upon increasing $V$
(see inset in Fig. \ref{fig15} ($b$)) \cite{Kumar2015}. The $T_c$ decreases with increasing $V$ for both $U=4$ and $U=6$. 
The trends for larger values of $U$ are similar to those for $U=6$.
The behavior of the system for different values of $V$ and $T$ is summarized in two phase diagrams (see Fig. \ref{fig15}($c$)-($d$)). 
For intermediate $U$, the SC order is destabilized with increasing temperature, giving way to a non-superconducting phase with finite spectral gap. 
This phase can be considered as a phase with trapped copper pairs. With further increase in $T$, the non-SC gapped phases evolves into a
pseudogapped phase (see Fig. \ref{fig15}($c$)). This phase suggests that the cooper pairs are not very robust and are at the verge of breaking into normal electrons.
For strong $U$, the non-SC gapped state is stable over wider region in $T$-$V$ space.

The QMC studies on AHM in two dimensions indicate an existence of a superconductor to insulator transition (SIT) upon increasing disorder strength
\cite{Scalettar1999}. The critical value of $V/t$ for $U/t = 4$ is found to lie between $3$ and $4$ for $n=0.86$ \cite{Scalettar1999}. The results 
obtained within our Monte Carlo method are consistent with the previous results. The critical value of disorder required for SIT increases with increasing $U$.
The pseudogap region, expands with increasing the strength of disorder for weak disorder, and reduces upon further increasing the disorder.

\section{Conclusions}

We have introduced an effective Hamiltonian based Monte-Carlo method for studying disordered AHM.
The method is inspired by the ideas presented by J. Hubbard in Phys. Rev. B {\bf 16}, 2626 (1979) in the context of repulsive Hubbard model. 
The interacting Hamiltonian is replaced by,
(i) an effective classical Hamiltonian that controls the fluctuations of the auxiliary fields, and (ii) a Hamiltonian describing
electrons in arbitrary potential arising due to the auxiliary field configurations. 
The parameters of the classical
Hamiltonian are determined from the behavior of energy variation about the BdG mean-field solutions. 
The results presented for
the disorder-free Hamiltonian are quantitatively close to
those reported in studies utilizing more sophisticated methods, such as, QMC, DMFT and SAF-MC. 
The effective Hamiltonian approach also provides additional insights into the behavior of the AHM.
The distribution of the auxiliary fields and their evolution with $U$ and $T$ can be used to make inference about the 
nature of the finite temperature phase transitions. We find that while the small $U$ (large $U$) limit is dominated by amplitude (phase)
fluctuations as expected in the BCS (BEC) scenario, both amplitude and phase fluctuations contribute significantly to the thermally induced 
suppression of superconductivity in the intermediate $U$ regime. The pseudogap phase exists in this regime just above $T_c$ 
when both amplitude and phase fluctuations are active. This agrees well with the recent experimental findings in NbN superconductors.
The advantage of the method lies in the fact that a purely classical Monte Carlo method can be employed to
generate auxiliary field configurations at finite temperatures.
Accessibility of large lattice sizes makes this a powerful method to study the effect of disorder on superconductivity.
To this end, we demonstrate that the method can indeed be used for disordered Hamiltonians. The parameters of the effective Hamiltonian become site- and bond-dependent
in the presence of quenched disorder. The effect of disorder is to enhance fluctuations in both the amplitude and phase variables. 
The observation of pseudogap in disordered s-wave superconductors is consistent with our inference that the pseudogap state is an indicator for
the presence of fluctuations in both phase and amplitude $\Delta_i$.
It will be an interesting future direction to explore the 
extension of our schemes to include superconducting phases with non-trivial order-parameter symmetries such as $d$-wave, $s^{++}/s^{+-}$-wave, etc.
The general idea of building an effective Hamiltonian by analysing the change in energy about the BdG mean-field state should work, provided one can
identify the relevant auxiliary fields that describe the low-energy fluctuations.

\section{Acknowledgments}
S.K. is grateful to T. V. Ramakrishnan for discussions and for pointing out reference \cite{Hubbard1979}. The calculations were performed using the 
High Performance Computing Facility at IISER Mohali. 
K.P. acknowledges support via UGC fellowship. S.K. acknowledges support from Department of Science and Technology (DST), India.

%


\end{document}